\documentclass[reprint,superscriptaddress,pra]{revtex4-2}

\usepackage{graphicx}
\usepackage{dcolumn}
\usepackage{bm}
\usepackage{graphicx}
\usepackage{bm}
\usepackage{pgfplots}
\usepackage{color}
\usepackage{algorithm}
\usepackage{url,array}
\usepackage{amsmath,amssymb,amsthm,dsfont,mathrsfs}
\usepackage{epstopdf}

\def\x{{\mathbf{x}}}
\def\y{{\mathbf{y}}}
\def\z{{\mathbf{z}}}
\def\M{{\mathbf{M}}}
\def\A{{\mathbf{A}}}
\def\I{{\mathbf{I}}}
\def\s{{\mathbf{s}}}

\def\c{{\mathbf{c}}}
\def\U{{\mathbf{U}}}
\def\u{{\mathbf{u}}}

\def\H{{\mathbf{H}}}
\def\v{{\mathbf{v}}}
\def\a{{\mathbf{a}}}
\def\b{{\mathbf{b}}}
\def\X{{\mathbf{X}}}
\def\Y{{\mathbf{Y}}}
\def\U{{\mathbf{U}}}
\def\Z{{\mathbf{Z}}}
\def\V{{\mathbf{V}}}
\def\T{{\mathbf{T}}}
\def\P{{\mathbf{P}}}
\begin{document}


\title{Arbitrarily-high-dimensional reconciliation via cross-rotation for\\ {continuous-variable quantum key distribution}}

\author{Jisheng Dai} 
\affiliation{College of Information Science and Technology, Donghua University, Shanghai 201620, China}
\author{Xue-Qin Jiang} \email{xqjiang@dhu.edu.cn}
\affiliation{College of Information Science and Technology, Donghua University, Shanghai 201620, China}
\affiliation{Hefei National Laboratory, Hefei 230088, China}

\author{Tao Wang}
\author{Peng Huang}
\author{Guihua Zeng}
\affiliation{State Key Laboratory of Advanced Optical Communication Systems and Networks, Shanghai Jiao Tong University, Shanghai 200240, China}
\affiliation{Hefei National Laboratory, Hefei 230088, China}


\begin{abstract}
Multidimensional rotation serves as a powerful tool for enhancing information reconciliation and extending the transmission distance in continuous-variable quantum key distribution (CV-QKD).  However, the lack of closed-form orthogonal transformations for high-dimensional rotations has limited the maximum reconciliation efficiency to
channels with 8 dimensions over the past decade.
This paper presents a cross-rotation scheme to overcome this limitation and enable reconciliation in arbitrarily high dimensions, {constrained to even multiples of $8$}. The key treatment involves reshaping the string vector into matrix form and applying orthogonal transformations to its columns and rows in a cross manner, thereby increasing the reconciliation dimension by one order per cross-rotation while significantly reducing the communication overhead over the classical channel.
A rigorous performance analysis is also presented from the perspective of achievable sum-rate.  Simulation results demonstrate that 64-dimensional cross-rotation nearly approaches the upper bound, making it a recommended choice for practical implementations.
\end{abstract}

\maketitle


\section{Introduction}


Information security is becoming increasingly crucial in today's digital society.
Quantum key distribution (QKD)  enables two distant parties (Alice and Bob) to establish an unconditionally secure key by exchanging quantum states, even in the presence of an eavesdropper, thereby ensuring complete privacy \cite{BEN84,RevModPhys.74.145,RevModPhys.84.621,Zhang2024APR}.
Continuous variable (CV)-QKD \cite{PhysRevLett.88.057902,PhysRevA.76.042305}
relies on the quadrature components of quantum states for encoding, making it highly compatible with current coherent optical communication systems \cite{ZhouC2019,PhysRevA.95.062330}. This compatibility allows CV-QKD to achieve higher key rates, particularly over short to medium distances, and benefits from more efficient modulation and decoding processes supported by existing technologies \cite{Yang2023EPJ}.
For CV-QKD protocols utilizing coherent states with Gaussian modulation, theoretical security proofs against various types of attacks have been established \cite{PhysRevLett.101.200504,PhysRevLett.114.070501,Matsuura2021,PhysRevLett.118.200501}.

Quantum transmission and post-processing are the two primary phases of a CV-QKD system. In the quantum transmission, Alice transmits quantum states to Bob via a quantum channel, where Bob measures them using either homodyne or heterodyne detection \cite{PhysRevA.76.042305}. The post-processing phase involves extracting secret keys through steps such as basis sifting, parameter estimation, information reconciliation, and privacy amplification \cite{RevModPhys.74.145,Yang2023EPJ}.
Information reconciliation, which corrects errors in raw keys through error-correcting codes, is particularly crucial as it directly affects the secret key rate (SKR) and the maximum secure transmission distance \cite{PhysRevApplied.8.044017}. Enhancing the efficiency of information reconciliation remains the primary challenge for improving CV-QKD system performance \cite{Laudenbach2018}.

Several reconciliation schemes have been proposed for CV-QKD, focusing primarily on slice reconciliation \cite{1266817} and multidimensional reconciliation \cite{PhysRevA.77.042325}.
Slice reconciliation involves quantization and error correction, while multidimensional reconciliation includes both transformation and error correction. Notably, multidimensional reconciliation applies a $d$-dimensional rotation to construct a virtual channel that closely approximates the binary input additive white Gaussian noise channel (BIAWGNC) from the physical Gaussian channel \cite{WOS:000432546600001}. Combined with robust error-correction codes, this approach significantly enhances reconciliation efficiency and secret key rates, especially in low signal-to-noise ratio conditions.
Numerous variants of multidimensional reconciliation have been discussed in the literature \cite{PhysRevA.84.062317,Jeong2022npj,Li2019QIP,PhysRevA.103.032603,PhysRevApplied.19.054084,ZhouC2019,Jiang2024}. For instance, \cite{PhysRevA.84.062317,Jeong2022npj} combined multidimensional reconciliation and low density parity-check (LDPC) codes to achieve high reconciliation efficiency. \cite{Li2019QIP} proposed a method for initial decoding message computation that eliminates the need for norm information from the encoder, while \cite{PhysRevA.103.032603} designed a decoding algorithm for the virtual channel where noise follows the Student's t-distribution. \cite{PhysRevApplied.19.054084} applied multidimensional reconciliation to non-Gaussian modulation protocols, further enhancing the performance of CV-QKD.
Clearly, all existing multidimensional methods aim to improve reconciliation efficiency by developing more effective encoding and decoding techniques.

Another potential strategy for improving reconciliation efficiency is to increase the dimensionality of reconciliation \cite{PhysRevA.77.042325}. However, the required closed-form rotation operator exists only in dimensions $d=1,2,4$ and $8$, restricting the use of multidimensional rotation to a maximum of eight dimensions \cite{WOS:000432546600001}. This limitation is reason most CV-QKD systems primarily use $8$-dimensional reconciliation ($d = 8$), as it offers the best performance among the currently available options.
Although Householder transformations can be used to recursively construct high-dimensional mapping matrices, as demonstrated in \cite{PhysRevA.84.062317, Zhao:22}, the communication overhead of transmitting numerous $d \times d$ mapping matrices over the classical channel becomes prohibitively intractable in practical implementations.

In this paper, we present a cross-rotation scheme to overcome the limitation of maximum $8$-dimensional closed-form rotation and extend it to arbitrarily high-dimensional reconciliation {(constrained to even multiples of $8$)}, with communication overhead scaling linearly with the number of cross-rotation operations. The key idea is to reshape the string vector into matrix form and apply orthogonal transformations to its columns and rows in a cross manner.
We also rigorously prove the performance improvement of the proposed high-dimensional reconciliation scheme from a signal processing perspective. Simulation results demonstrate the superiority of the proposed algorithm over state-of-the-art counterparts.



%


\section{Review of multidimensional reconciliation}

After transmitting quantum states, sifting data, and estimating parameters in a CV-QKD system, Alice and Bob will generate correlated Gaussian sequences of length $N$, denoted by $\x$ and $\y$ at the sides of Alice and Bob, respectively.
Let $\z$ denote the  i.i.d. Gaussian quantum channel noise with an elemental zero mean and variance $\sigma^2$, then $\y$ can be written as:
\begin{align}
\y=\x+ \z,\label{eq-yxz}
\end{align}
and $p(\y|\x)=\mathcal{N}(\y|\x, \sigma^2\I)$.
Due to the presence of quantum channel noise $\z$, $\y$ cannot be perfectly matched with $\x$, potentially resulting in significant degradation of the distilled secret keys.
Information reconciliation is a vital step in correcting errors to establish shared secret keys between two communicating parties while revealing as little information as possible about the secret keys.
There are two distinct information reconciliation methods: direct reconciliation (where Bob corrects bits with Alice's assistance) and reverse reconciliation (where Alice corrects bits with Bob's assistance) \cite{PhysRevLett.88.057902}.
Reverse reconciliation is known to outperform direct reconciliation in terms of transmission distance and maintains a stable SKR \cite{Grosshans2003,PhysRevA.76.052301}.

Among the various schemes proposed for the reverse information reconciliation of Gaussian symbols, {the multidimensional  reconciliation is highly efficient for long-distance CV-QKD with a low  signal-to-noise ratio (SNR) \cite{PhysRevA.77.042325,PhysRevA.84.062317,WOS:000432546600001}.}
This approach transforms the channel between Alice and Bob into a virtual BIAWGNC, which enables the use of efficient binary codes. Multidimensional reverse reconciliation is divided into two phases: 1) Bob performs multidimensional processing on his received quantum signal $\y$ and transmits the necessary information to Alice via the classical channel; and 2) Alice corrects errors and extracts the secret keys using the multidimensional information received from Bob. Specifically,
\begin{itemize}
  \item For $d$-dimensional reverse reconciliation, Bob first divides $\y$ into $G=N/d $ subsequences,
denoted as $\y=[\y_1^{\mathrm{T}},\y_2^{\mathrm{T}},\ldots, \y_G^{\mathrm{T}} ]^{\mathrm{T}}$, where $\y_{g}\in \mathbb{R}^d$ and $(\cdot)^{\mathrm{T}}$ denotes vector/matrix transpose. He then generates
an $(N, K)$ LDPC codeword
$\c=[\c_1^{\mathrm{T}}, \c_2^{\mathrm{T}},\ldots, \c_G^{\mathrm{T}} ]^{\mathrm{T}}$
and calculates its syndrome as $\s=\H\c$ , where $\c_g$ is of length $d$ with each element $c_{i,g}\in \{0,1\}$ and $\H \in \mathbb{R}^{M\times N}$ denotes the parity-check matrix with $M=N-K$.
After that Bob transforms each random bit string $\c_g$ into a $d$-dimensional spherical vector as:
\begin{align}
\u_g=\left[ \frac{ (-1)^{c_{1,g}}}{\sqrt{d}} , \frac{ (-1)^{c_{2,g}}}{\sqrt{d}},\ldots, \frac{ (-1)^{c_{d,g}}}{\sqrt{d}}     \right]^{\mathrm{T}}.
\end{align}
If $d=1, 2, 4$ or $8$,  there always exist a closed-form orthogonal mapping matrix $\M(\y_g, \u_g)\triangleq\sum_{i=1}^d \alpha_i(\y_g, \u_g) \A_i $ such that \cite{PhysRevA.77.042325}:
\begin{align}\label{eq-mygug}
\M(\y_g, \u_g) \frac{\y_g}{\| \y_g\|} = \u_g,
\end{align}
or, equivalently,
\begin{align}\label{eq-mygug2}
\M(\y_g, \u_g) \y_g = \| \y_g\| \u_g,
\end{align}
where $\alpha_i(\y_g, \u_g)= \u_g^{\mathrm{T}}\A_i \y_g /\| \y_g\|$ denotes the $i$-th mapping coefficient  and $\A_i\in\mathbb{R}^{d\times d}$ denotes a fixed orthogonal matrix (defined in Appendix of \cite{PhysRevA.77.042325}). It is straightforward to show that $\M(\y_g, \u_g)$ is a unitary matrix and is completely determined by the mapping coefficients $\{\alpha_i(\y_g, \u_g)\}_{i=1}^d$. Finally, Bob sends the mapping coefficients, the syndrome $\s$ and the vector norms $\{\|\y_g\|\}_{g=1}^G$ to Alice over the classical channel to assist her in the distillation of the secret keys.

  \item   Alice initially constructs the  mapping matrix $\M(\y_g, \u_g)$ using the received mapping coefficients, and
  divides $\x$ into $G$ subsequences, leading to:
\begin{align}\label{eq-xyzg}
\x_g = \y_g - \z_g,
\end{align}
where $\x_g$ and $\z_g$ are similarly defined as $\y_g$.
Left-multiplying (\ref{eq-xyzg}) with $\M(\y_g, \u_g)$ yields:
\begin{align}\label{eq-Mxuz}
\M(\y_g, \u_g)\x_g = \|\y_g\|\u_g  + \z'_g,
\end{align}
where the equality directly follows from (\ref{eq-mygug2}) and $\z'_g\triangleq -\M(\y_g, \u_g)\z_g $ has the same i.i.d. Gaussian distribution as $\z_g$ due to the  orthogonal invariance property of Gaussian random variables \cite{Harpaz2005}. Given that $\|\y_g\|$ is known to Alice via the classical channel, (\ref{eq-Mxuz}) can be rewritten as:
\begin{align}\label{eq-Mxyuz}
\underbrace{\frac{\M(\y_g, \u_g)}{ \|\y_g\|}\x_g}_{\triangleq \v_g} =\u_g  + \frac{\z'_g}{ \|\y_g\|}.
\end{align}
Finally, Alice attempts to decode an estimate $\hat{\c}$ of the codeword $\c$ generated by Bob using the sum-product algorithm with (\ref{eq-Mxyuz}). LDPC decoding is successful if $\H\hat{\c}=\s$.
\end{itemize}

It is well known that increasing the dimensionality of the reconciliation process can significantly improve reconciliation efficiency \cite{PhysRevA.77.042325,PhysRevA.84.062317}. Over the past decade, due to the lack of a closed-form orthogonal transformation for high-dimensional rotations ($d > 8$), the highest reconciliation efficiency has remained at an $8$-dimensional channel (i.e., $d = 8$).
While Householder transformations were employed to recursively construct high-dimensional mapping matrices in \cite{PhysRevA.84.062317, Zhao:22}, the communication overhead of transmitting the $d \times d$ mapping matrix over the classical channel becomes prohibitively large.
These limitations have prompted us to explore new methods for achieving higher-dimensional rotations while maintaining a low communication overhead.


\section{Proposed cross-rotation scheme}
In this section, we present a cross-rotation scheme that significantly extends reconciliation dimensionality.
To simplify the presentation, we first focus on the cross-rotation for $64$-dimensional reconciliation (i.e., $d = 64$),
and then illustrate how iterative application of the cross-rotation scheme can construct arbitrarily high-dimensional reconciliation.
The following subsections begin by detailing the cross encoding and decoding processes on both parties' sides, followed by a discussion on how to derive the initial log-likelihood ratio (LLR) message for the sum-product decoding. Afterward, we provide a performance analysis of the proposed method and demonstrate that there is no information leakage.


\subsection{64-dimensional encoding on Bob's side}

Bob first reshapes each  $\y_g \in \mathbb{R}^{64}$ into a matrix $\Y_g \in \mathbb{R}^{8 \times 8}$, and applies the same reshaping to $\u_g\in \mathbb{R}^{64} $ to form $\U_g\in \mathbb{R}^{8 \times 8}$.
{It is worth noting that $\U_g$ is solely used for aligning the norms of each row of ${\Y}_g$ (which will be shown in (\ref{eq-UYG})) and shall be discarded once applied.}
Specifically, let $\Y_{:l,g}$ and $\U_{:l,g}$ denote the $l$-th columns of  $\Y_g$ and $\U_g$, respectively.
Then, each column vector  $\Y_{:l,g}$  can be orthogonally transformed into:
\begin{align}
\M(\Y_{:l,g}, \U_{:l,g}) \cdot  \Y_{:l,g}   =  \| \Y_{:l,g} \| \cdot\U_{:l,g}. \label{eq-MYUY}
\end{align}
By arranging these rotated column vectors in matrix form, he obtains:
\begin{align}
 &\begin{bmatrix} \M_{1,g} \cdot \Y_{:1,g}  ,~ \M_{2,g}  \cdot \Y_{:2,g} ,~ \ldots ,~  \M_{8,g} \cdot \Y_{:8,g}  \end{bmatrix}\notag\\
 =&\begin{bmatrix}  \| \Y_{:1,g} \| \cdot\U_{:1,g} ,~  \| \Y_{:2,g}\| \cdot\U_{:2,g} ,~ \ldots ,~   \| \Y_{:8,g} \| \cdot\U_{:8,g}  \end{bmatrix}\notag\\
=& \underbrace{ \U_g \cdot \begin{bmatrix}   \| \Y_{:1,g} \| & & &\\
                           & \| \Y_{:2,g} \|& & \\
                           & &  \ddots  &  \\
                           & &    &   \| \Y_{:8,g} \| \end{bmatrix}}_{\triangleq \tilde{\Y}_{g}  },   \label{eq-UYG}
\end{align}
where $\M_{l,g}$ is short for $\M(\Y_{:l,g}, \U_{:l,g}) $.

\begin{figure}
  \centering
    \includegraphics[width=8.0cm]{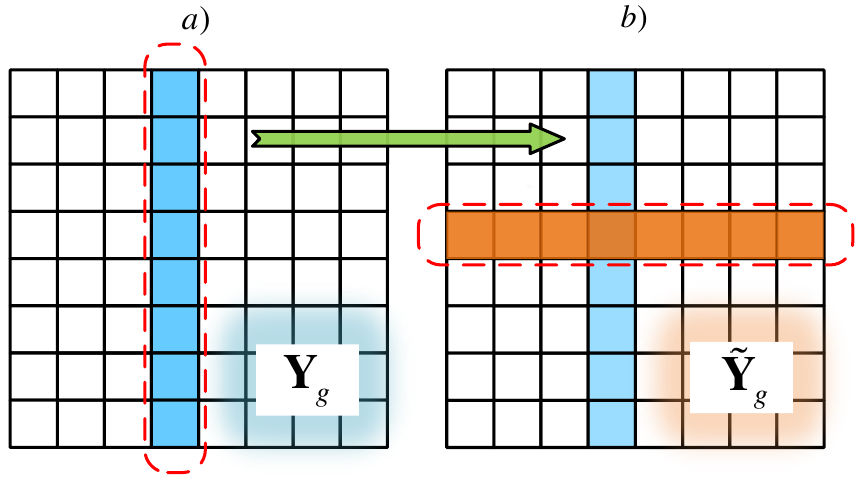}\\
  \caption{Illustration of  cross-rotation scheme. a) column rotation for ${\Y}_g$ ; and b) row rotation for $\tilde{\Y}_g$.}\label{fig1}
\end{figure}

The key idea of the proposed cross-rotation scheme is to apply the orthogonal transformation to $\tilde{\Y_g}$ once again, but this time, it is applied to the rows (as shown in Fig.~\ref{fig1}).
{Bob will generate a new spherical matrix $\tilde{\U}_g$ similarly to $\U_g$, which will be used to distill the secret key between the two parties.}
Specifically, for the $l$-th row vector of $\tilde{\Y}_g$, it will be orthogonally transformed as follows:
\begin{align}
\tilde{\Y}_{l:,g} \cdot  \M^{\mathrm{T}}(\tilde{\Y}_{l:,g}^{\mathrm{T}} , \tilde{\U}_{l:,g}^{\mathrm{T}})
=& \sqrt{ \sum\nolimits_{l=1}^8  \frac{ \| \Y_{:l,g} \|^2}{8}   }  \cdot  \tilde{\U}_{l:,g} \notag\\
=&  \frac{ \| \Y_{g} \|_{\mathrm{F}}}{\sqrt{8}} \cdot \tilde{\U}_{l:,g}, \label{eq-YFU8}
\end{align}
where $\tilde{\Y}_{l:,g}$ and $\tilde{\U}_{l:,g}$ denote the $l$-th rows of  $\tilde{\Y}_g$ and $\tilde{\U}_g$, respectively, and $\|\cdot\|_{\mathrm{F}}$ denotes the Frobenius norm of a matrix.
Stacking these rotated rows back into matrix form yields:
\begin{align}
\begin{bmatrix}
\tilde{\Y}_{1:,g} \cdot \tilde{\M}_{1,g}^{\mathrm{T}}\\
\tilde{\Y}_{2:,g} \cdot \tilde{\M}_{2,g}^{\mathrm{T}}\\
\vdots\\
\tilde{\Y}_{8:,g} \cdot \tilde{\M}_{8,g}^{\mathrm{T}}
\end{bmatrix}=
\begin{bmatrix}
\frac{ \| \Y_{g} \|_{\mathrm{F}}}{\sqrt{8}} \cdot \tilde{\U}_{1:,g}\\
\frac{ \| \Y_{g} \|_{\mathrm{F}}}{\sqrt{8}} \cdot \tilde{\U}_{2:,g}\\
\vdots\\
\frac{ \| \Y_{g} \|_{\mathrm{F}}}{\sqrt{8}} \cdot \tilde{\U}_{8:,g}
\end{bmatrix}
=\frac{ \| \Y_{g} \|_{\mathrm{F}}}{\sqrt{8}} \cdot  \tilde{\U}_g, \label{eq-ygug}
\end{align}
where $\tilde{\M}_{l,g}$ is short for $\M(\tilde{\Y}_{l:,g}^{\mathrm{T}} , \tilde{\U}_{l:,g}^{\mathrm{T}}) $.
Finally, Bob sends the mapping coefficients for $\{\M_{l,g},\forall l,g\}$ and $\{\tilde{\M}_{l,g},\forall l,g\}$,
along with the syndromes and the matrix norms $\{\|\Y_g\|,\forall g\}$, to Alice over the classical channel to help her distill the secret keys contained in $\tilde{\U}_g$.
Clearly, the communication overhead over the classical channel is $\mathcal{O}(2N)$ for the proposed 64-dimensional reconciliation.

\subsection{64-dimensional decoding on Alice's side}

Alice initially constructs the mapping matrices $\{\M_{l,g},\forall l,g\}$ and $\{\tilde{\M}_{l,g},\forall l,g\}$ using the received coefficients. With (\ref{eq-xyzg}), $\x_g \in \mathbb{R}^{64}$ can be reshaped into:
\begin{align}
\X_g = \Y_g  - \Z_g,
\end{align}
where both $\X_g \in \mathbb{R}^{8\times 8}$ and $\Z_g \in \mathbb{R}^{8\times 8}$  are similarly defined as $\Y_g$.
Left-multiplying the $l$-th column of $\X_g$ (denoted as $\X_{:l,g}$) by $\M_{l,g}$ gives:
\begin{align}
&\M_{l,g}\cdot \X_{:l,g}=  \M_{l,g} \cdot \Y_{:l,g}  +  \Z'_{:l,g}=  \| \Y_{:l,g} \| \cdot\U_{:l,g}   +  \Z'_{:l,g},
\end{align}
where $\Z'_{:l,g}\triangleq -\M_{l,g} \Z_{:l,g}$ has the same i.i.d. Gaussian distribution as $\Z_{:l,g}$.
Arranging these rotated columns into matrix form yields:
\begin{align}
 \tilde{\X}_{g} = \tilde{\Y}_{g}   +  \Z'_g,
\end{align}
where $\tilde{\X}_{g}\triangleq [ \M_{1,g} \cdot \X_{:1,g}  ,~ \M_{2,g}  \cdot \X_{:2,g} ,~ \ldots ,~  \M_{8,g} \cdot \X_{:8,g} ]$ and $\tilde{\Y}_{g}$ is defined in (\ref{eq-UYG}).

Right-multiplying the $l$-th row of $\tilde{\X}_{g}$ (denoted as $\tilde{\X}_{l:,g}$) by $\tilde{\M}_{l,g}$ and stacking the result back into matrix form results in:
\begin{align}\label{eq-BYUZ}
\begin{bmatrix}
\tilde{\X}_{1:,g} \cdot \tilde{\M}_{1,g}^{\mathrm{T}}\\
\tilde{\X}_{2:,g} \cdot \tilde{\M}_{2,g}^{\mathrm{T}}\\
\vdots\\
\tilde{\X}_{8:,g} \cdot \tilde{\M}_{8,g}^{\mathrm{T}}
\end{bmatrix}=
\frac{ \| \Y_{g} \|_{\mathrm{F}}}{\sqrt{8}} \cdot  \tilde{\U}_g  +  \tilde{\Z}_g,
\end{align}
where $\tilde{\Z}_g$  also has the same i.i.d. Gaussian distribution as $\Z_g$, and the equality directly follows from (\ref{eq-ygug}).
Given that $\|\Y_{g}\|$ is known to Alice via the classical channel, (\ref{eq-BYUZ}) can be rewritten as:
\begin{align}\label{eq-1yxmuz}
\underbrace{\frac{\sqrt{8}}{\| \Y_{g} \|_{\mathrm{F}}} \begin{bmatrix}
\tilde{\X}_{1:,g} \cdot \tilde{\M}_{1,g}^{\mathrm{T}}\\
\tilde{\X}_{2:,g} \cdot \tilde{\M}_{2,g}^{\mathrm{T}}\\
\vdots\\
\tilde{\X}_{8:,g} \cdot \tilde{\M}_{8,g}^{\mathrm{T}}
\end{bmatrix}}_{\triangleq\tilde{\V}_g}=    \tilde{\U}_g  +  \frac{\sqrt{8}}{\| \Y_{g} \|_{\mathrm{F}}} \tilde{\Z}_g.
\end{align}

Vectorizing both sides of (\ref{eq-1yxmuz}) gives:
\begin{align}\label{eq-vectorgug}
\tilde{\v}_g= \tilde{\u}_g  +  \frac{\sqrt{8}}{\| \Y_{g} \|_{\mathrm{F}}}\tilde{\z}_g,
\end{align}
where $\tilde{\v}_g$, $\tilde{\u}_g$ and $\tilde{\z}_g$ denote the vectorizations of  $\tilde{\V}_g$, $\tilde{\U}_g$ and $\tilde{\Z}_g$, respectively. Once the LLR message is derived under the data model (\ref{eq-vectorgug}), LDPC decoding can then be directly applied to correct the errors. Although we present the rotation sequence as column rotation followed by row rotation, it can also be justified as row rotation first, followed by column rotation.

\subsection{Initial LLR message}

In this subsection, we discuss the derivation of the initial LLR message for sum-product LDPC decoding under the cross-rotation model.
According to (\ref{eq-vectorgug}), the elemental probability density function of $\tilde{\v}_g$, conditioned on $\tilde{\u}_g$, is given as:
\begin{align}\label{eq-pvu}
p(\tilde{v}_{i,g} | \tilde{u}_{i,g} )=  \mathcal{N}\Big(\tilde{v}_{i,g} \Big|  \tilde{u}_{i,g},\frac{8\sigma^2}{\| \Y_{g} \|_{\mathrm{F}}^2}\Big),
\end{align}
where $\tilde{v}_{i,g}$ and $\tilde{u}_{i,g}$ denote the $i$-th elements of $\tilde{\v}_g$ and $\tilde{\u}_g$, respectively.
Using Bayes' theorem, the elemental posteriori probability is given as:
\begin{align}\label{eq-puv}
p(\tilde{u}_{i,g} | \tilde{v}_{i,g} ) = \frac{p(\tilde{v}_{i,g} | \tilde{u}_{i,g} ) p(\tilde{u}_{i,g} )}{p(\tilde{v}_{i,g} )}.
\end{align}
Then, the initial LLR message of $\tilde{u}_{i,g} $ can be calculated as:
\begin{align}
\mathrm{LLR}^0_{i,g}\triangleq & \ln\frac{p(\tilde{u}_{i,g}=\frac{1}{\sqrt{8}} | \tilde{v}_{i,g} )}{p(\tilde{u}_{i,g}=\frac{-1}{\sqrt{8}} | \tilde{v}_{i,g} )}\notag\\
=& \ln \frac{p(\tilde{v}_{i,g} | \tilde{u}_{i,g}=\frac{1}{\sqrt{8}}  )}{p(\tilde{v}_{i,g} | \tilde{u}_{i,g}=\frac{-1}{\sqrt{8}}  )}\notag\\
=& \frac{ \| \Y_{g} \|_{\mathrm{F}}^2 \cdot \tilde{v}_{i,g}   }{ 8 \sqrt{2}\sigma^2     },
\end{align}
where the last two equalities follow (\ref{eq-puv}) and (\ref{eq-pvu}), respectively.

\subsection{Performance analysis}

In this subsection, we analyze the performance improvement of multidimensional reverse reconciliation for secret key distillation from a signal processing perspective. Specifically, we prove that the proposed $64$-dimensional cross-reconciliation outperforms the traditional $8$-dimensional reconciliation in terms of achievable sum-rate.
Note that although the following proof applies to arbitrary high-dimensional cross-reconciliation,
we focus on the $64$-dimensional case for simplicity.

We proceed with the performance analysis of multidimensional reconciliation using majorization, and some related fundamental definitions and lemma are provided in Appendix~A.
Recall that $\tilde{\v}_g$  (given in (\ref{eq-vectorgug})) has a size of $64\times 1$, and the achievable data rate  $\tilde{R}_{i,g}$ for the $i$-th symbol of $\tilde{\u}_g$ is given by \cite{Cover2006}:
\begin{align}
\tilde{R}_{i,g}=& \frac{1}{2}\log_2\left( 1+  \frac{1/8}{8\sigma^2/\|\Y_g \|_\mathrm{F}^2}   \right)\notag\\
=& \frac{1}{2}\log_2\left( 1+  \frac{\|\Y_g \|_\mathrm{F}^2}{64\sigma^2}   \right),~i =1,2,\ldots, 64.
\end{align}
Therefore, the sum-rate for the $64$-dimensional cross-rotation (across all $64$ symbols), as given in (\ref{eq-vectorgug}), is:
\begin{align}\label{eq-rg64}
\tilde{R}_g= \sum_{i=1}^{64} \tilde{R}_{i,g}   = 32\cdot \log_2\left( 1+  \frac{\|\Y_g \|_\mathrm{F}^2}{64\sigma^2}   \right).
\end{align}

On the other hand, when the traditional $8$-dimension rotation is used, the achievable data rate for $\Y_{:l,g}$ (across $8$ symbols) is:
\begin{align}
R_{l,g}= 4 \cdot \log_2\left( 1+  \frac{\|\Y_{:l,g} \|_\mathrm{F}^2}{8\sigma^2}   \right),~l=1,2,\ldots, 8,
\end{align}
and the sum-rate for $\Y_g$ (across all $64$ symbols) is:
\begin{align}\label{eq-rg8}
R_g= \sum_{l=1}^{8}R_{l,g} = 4 \cdot \sum_{l=1}^8 \log_2\left( 1+  \frac{\|\Y_{:l,g} \|_\mathrm{F}^2}{8\sigma^2}   \right).
\end{align}

Define  $f(\a)=\sum_{i=1}^{8}\log_2(1+  a_i) $, where $\a=[a_1, a_2, \ldots, a_{8}]^{\mathrm{T}}$.
Then, (\ref{eq-rg64}) and (\ref{eq-rg8}) can be rewritten as:
\begin{align}
\tilde{R}_g= 4 \cdot f(\tilde{\bm\gamma}),\\
R_g=  4 \cdot f(\bm\gamma),
\end{align}
respectively, with
\begin{align*}
\tilde{\bm\gamma}=&  \Big[\frac{1}{64\sigma^2}\|\Y_g \|_\mathrm{F}^2, \frac{1}{64\sigma^2}\|\Y_g \|_\mathrm{F}^2, \ldots, \frac{1}{64\sigma^2}\|\Y_g \|_\mathrm{F}^2\Big]^{\mathrm{T}}\in \mathbb{R}^8,\\
\bm\gamma=&  \Big[\frac{1}{8\sigma^2} \|\Y_{:1,g} \|_\mathrm{F}^2,   \frac{1}{8\sigma^2} \|\Y_{:2,g} \|_\mathrm{F}^2,  \ldots  ,  \frac{1}{8\sigma^2}\|\Y_{:8,g} \|_\mathrm{F}^2\Big]^{\mathrm{T}}\in \mathbb{R}^8.
\end{align*}

According to Definition~1 in Appendix~A, it is easy to verify that  $\tilde{\bm\gamma}  \prec \bm\gamma  $. Since $f(\bm\gamma)$ is Schur-concave (see Lemma~3),  we can conclude that:
\begin{align}
\tilde{R}_g= 4 \cdot f(\tilde{\bm\gamma}) \ge   4 \cdot f(\bm\gamma)   =  R_g,
\end{align}
which follows directly from Definition~2 in Appendix~A.
Hence, the proposed cross-rotation achieves a higher sum-rate than the traditional approach, indicating better performance in various SNR scenarios.

\subsection{No information leakage}

In this subsection, we demonstrate that revealing the mapping coefficients does not disclose any information under the proposed cross-rotation scheme. For ease of presentation, assume $\y\in \mathbb{R}^d$ has unit norm, and denote the mapping coefficients for $\u\in \mathbb{R}^d$ and $\tilde{\u}\in \mathbb{R}^d$ as $\bm\alpha=[\alpha_1, \alpha_2, \ldots,\alpha_d]^{\mathrm{T}}$
and $\bm\beta=[\beta_1, \beta_2, \ldots,\beta_d]^{\mathrm{T}}$, respectively, where $\alpha_j=\u^{\mathrm{T}}\A_j\y$ and
$\beta_j=\tilde{\u}^{\mathrm{T}}\A_j\u$, $j=1,2,\ldots, d$.
Since mutual information is zero for statistically independent variables (see Appendix~B), proving that revealing $\bm\alpha$ and $\bm\beta$ does not leak any information about $\u$, $\tilde{\u}$, or $\y$ is equivalent to proving the following equalities:
\begin{align}
p(\u|\bm\alpha, \bm\beta) &= p(\u), \label{G1}\\
p(\tilde{\u}|\bm\alpha, \bm\beta) &= p(\tilde{\u}), \label{G2}\\
p(\y|\bm\alpha, \bm\beta) &= p(\y),\label{G3}
\end{align}
respectively.

Based on the definitions of $\alpha_j$s, the vector $\bm\alpha$ can be written as:
\begin{align}
\bm\alpha
=
\underbrace{\begin{bmatrix}
\u^{\mathrm{T}}\A_1\\
\u^{\mathrm{T}}\A_2\\
\vdots\\
\u^{\mathrm{T}}\A_d
\end{bmatrix}}_{\triangleq \M(\u) }\y \label{eq-Ba},
\end{align}
or, equivalently,
\begin{align}
\y=  \big(\M(\u)\big)^{-1}\bm\alpha=  \big(\M(\u)\big)^{\mathrm{T}}\bm\alpha  = \underbrace{\left(\sum_{j=1}^d \alpha_j\A_j^{\mathrm{T}}\right)}_{ \triangleq \M(\bm\alpha)  }  \u  \label{eq-By},
\end{align}
where the second equality follows from the fact that $\M(\u)$ is a unitary matrix.
Similarly, we have
\begin{align}
\bm\beta
=
\begin{bmatrix}
\tilde{\u}^{\mathrm{T}}\A_1\\
\tilde{\u}^{\mathrm{T}}\A_2\\
\vdots\\
\tilde{\u}^{\mathrm{T}}\A_d
\end{bmatrix} \u
=
\underbrace{\begin{bmatrix}
\u^{\mathrm{T}}\A_1^{\mathrm{T}}\\
\u^{\mathrm{T}}\A_2^{\mathrm{T}}\\
\vdots\\
\u^{\mathrm{T}}\A_d^{\mathrm{T}}
\end{bmatrix}}_{  \triangleq \tilde{\M}(\u)  } \tilde{\u} \label{eq-Bb},
\end{align}
or, equivalently,
\begin{align}
\tilde{\u}=   \underbrace{\left(\sum_{j=1}^d \beta_j\A_j\right)}_{ \triangleq \tilde{\M}(\bm\beta)  } \u \label{eq-Bu2}.
\end{align}
With (\ref{eq-Ba})--(\ref{eq-Bu2}), we are now ready to prove (\ref{G1})--(\ref{G3}) one-by-one using the sum-product message passing technique \cite{910572}.

\begin{proof} We begin by proving that (\ref{G1}) holds.
The joint probability density function (PDF) of $\{  \y,\u,\tilde{\u} ,\bm\alpha, \bm\beta\}$  can be expressed as
\begin{align}
p(\y,\u,\tilde{\u} , \bm\alpha, \bm\beta )= p(\bm\alpha| \y, \u) p(\bm\beta| \tilde{\u}, \u)  p(\y) p(\u) p(\tilde{\u}).
\end{align}
The corresponding factor graph is shown in Fig.~\ref{fig7}, where blank circles and black squares represent variable
nodes and factor nodes, respectively. Note that $p(\y)$ follows a sphere uniform distribution (denoted as $\mathrm{Uniform}_S(\y)$), while both $p(\u)$ and $p(\tilde{\u})$ follow discrete uniform distributions (denoted as $\mathrm{Uniform}_D(\u)$ and $\mathrm{Uniform}_D(\tilde{\u})$, respectively). The PDF at Node $\mathcal{YU}$ is given by $p(\bm\alpha| \y, \u)=\delta\big(\bm\alpha- \M(\u) \y \big)$ due to (\ref{eq-Ba}), where $\delta(\cdot)$ is the Dirac delta function. Similarly, the PDF at Node $\mathcal{U}\tilde{\mathcal{U}}$ is $p(\bm\beta| \tilde{\u}, \u)=\delta\big(\bm\beta- \tilde{\M}(\u) \tilde{\u} \big) $ due to (\ref{eq-Bb}).

\begin{figure}
\center
\includegraphics[width=5.5cm]{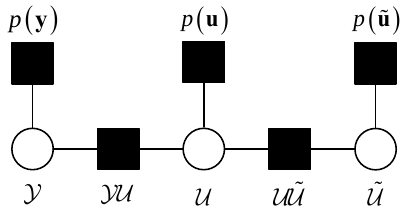}\\
\caption{Illustrations of the factor graph for message passing.}\label{fig7}
\end{figure}

Let $\Delta_{a\rightarrow b}(\cdot)$ denote the message passed from Node $a$ to Node $b$. According to the sum-product update rule \cite{910572}, the message passed from Node $\mathcal{YU}$ to Node $\mathcal{U}$ can be written as:
\begin{align}
\Delta_{\mathcal{YU}\rightarrow \mathcal{U}}(\u) \propto & \int \Delta_{\mathcal{Y}\rightarrow \mathcal{YU}}(\y)  \cdot p(\bm\alpha| \y, \u) d\y \label{eq-BYUU1}\\
\propto &  \int \mathrm{Uniform}_S(\y) \cdot \delta\big(\bm\alpha- \M(\u) \y \big)  d\y \label{eq-BYUU2}\\
\propto &  \int \mathrm{Uniform}_S(\y) \cdot \delta\big(\y- \M(\bm\alpha) \u \big)  d\y \label{eq-BYUU3}\\
\propto &~  \mathrm{Uniform}_C(\u)  \label{eq-BYUU4},
\end{align}
where $\propto$ denotes equality up to a proportional constant, and $\mathrm{Uniform}_C(\cdot)$ denotes a continuous uniform distribution.
Here, (\ref{eq-BYUU3}) follows from the equivalence between (\ref{eq-Ba}) and (\ref{eq-By}), and (\ref{eq-BYUU4}) follows from the fact that the expression of $\mathrm{Uniform}_S(\y)$ is noninformative with respect to $\y$.
Similarly, we obtain
\begin{align}
\Delta_{\mathcal{U}\tilde{\mathcal{U}}\rightarrow \mathcal{U}}(\u) \propto & \int \Delta_{\tilde{\mathcal{U}}\rightarrow \mathcal{U}\tilde{\mathcal{U}}}(\tilde{\u})  \cdot p(\bm\beta| \tilde{\u}, \u) d\tilde{\u} \notag\\
\propto &  \int \mathrm{Uniform}_D(\tilde{\u}) \cdot \delta\big(\bm\beta- \tilde{\M}(\u) \tilde{\u} \big)   d\tilde{\u} \notag\\
\propto &  \int \mathrm{Uniform}_D(\tilde{\u}) \cdot \delta\big(\tilde{\u}- \tilde{\M}(\bm\beta) \u \big)   d\tilde{\u} \notag\\
\propto &~  \mathrm{Uniform}_C(\u). \label{eq-BUtUU}
\end{align}
With (\ref{eq-BYUU4}) and (\ref{eq-BUtUU}), the sum-product belief at Node $\mathcal{U}$ can be calculated by:
\begin{align}
b_{sp}(\u) \propto &  \Delta_{\mathcal{YU}\rightarrow \mathcal{U}}(\u) \cdot
\Delta_{\mathcal{U}\tilde{\mathcal{U}}\rightarrow \mathcal{U}}(\u) \cdot p(\u) \notag\\
\propto &  \mathrm{Uniform}_C(\u)\cdot  \mathrm{Uniform}_C(\u) \cdot  \mathrm{Uniform}_D(\u) \notag\\
\propto &   \mathrm{Uniform}_D(\u).
\end{align}
As shown in \cite{910572},  the sum-product belief $b_{sp}(\u)$ corresponds to the posterior distribution $p(\u|\bm\alpha, \bm\beta)$. Therefore, we arrive at the desired conclusion $p(\u|\bm\alpha, \bm\beta) = p(\u)$.

Then, we show that (\ref{G2}) holds in the same manner.
The message passed from Node  $\mathcal{U}\tilde{\mathcal{U}}$  to Node $\tilde{\mathcal{U}}$ is:
\begin{align}
&\Delta_{\mathcal{U}\tilde{\mathcal{U}}\rightarrow \tilde{\mathcal{U}}}(\tilde{\u}) \notag\\
\propto & \int \Delta_{\mathcal{U}\rightarrow \mathcal{U}\tilde{\mathcal{U}}}(\u)  \cdot p(\bm\beta| \tilde{\u}, \u) d\u \notag\\
\propto & \int \Delta_{\mathcal{YU}\rightarrow \mathcal{U}}(\u)  \cdot  p(\u)  \cdot p(\bm\beta| \tilde{\u}, \u) d\u \notag\\
\propto & \int  \mathrm{Uniform}_C(\u)    \cdot \mathrm{Uniform}_D(\u)   \cdot \delta\big(\tilde{\u}- \tilde{\M}(\bm\beta) \u \big)   d\u \notag\\
\propto &~ \mathrm{Uniform}_C(\tilde{\u}).
\end{align}
Therefore, the sum-product belief at Node $\tilde{\mathcal{U}}$ can be calculated by:
\begin{align}
b_{sp}(\tilde{\u}) \propto &  \Delta_{\mathcal{U}\tilde{\mathcal{U}} \rightarrow \tilde{\mathcal{U}}}(\tilde{\u}) \cdot p(\tilde{\u}) \notag\\
\propto &  \mathrm{Uniform}_C(\tilde{\u}) \cdot  \mathrm{Uniform}_D(\bar{\u}) \notag\\
\propto &   \mathrm{Uniform}_D(\tilde{\u}).
\end{align}
Since the sum-product belief $b_{sp}(\tilde{\u})$ corresponds to the posterior distribution $p(\tilde{\u}|\bm\alpha, \bm\beta)$, it follows that (\ref{G2}) holds.

Finally, the sum-product belief at Node $\mathcal{Y}$ can be calculated similarly as:
\begin{align}
b_{sp}(\y) \propto &  \Delta_{\mathcal{YU} \rightarrow \mathcal{Y}}(\u) \cdot p(\y) \notag\\
\propto & \Bigg(\int  \Delta_{\mathcal{U}\tilde{\mathcal{U}}\rightarrow \mathcal{U}}(\u) \cdot p(\u) \cdot p(\bm\alpha| \y, \u) d\u \Bigg)  \cdot p(\y) \notag\\
\propto &~ \mathrm{Uniform}_S(\y)
\end{align}
which confirms that (\ref{G3}) holds.
\end{proof}



\subsection{Extension to arbitrarily high-dimensional reconciliation}

The significant contribution of the above cross-rotation scheme is that it breaks the limitation of the maximum $8$-dimensional rotation by enabling a $64$-dimensional rotation. In fact, iteratively applying the cross rotation can be used to achieve even higher-dimensional rotations. For instance, by vectorizing both sides of (\ref{eq-ygug}), we obtain:
\begin{align}
\vec{\y}_g =  \frac{ \| \Y_{g} \|_{\mathrm{F}}}{\sqrt{8}}  \tilde{\u}_g,
\end{align}
where $\vec{\y}_g$ is short for the vectorization of the left side of (\ref{eq-ygug}).
By arranging $\vec{\y}_g$ and $\tilde{\u}_g$ for $g=1,2,\ldots, 8$ into matrices, we have
\begin{align}
&\begin{bmatrix}
\vec{\y}_1, \vec{\y}_2, \ldots, \vec{\y}_8
\end{bmatrix}\notag\\
=&
\begin{bmatrix}
\tilde{\u}_1, \tilde{\u}_2, \ldots, \tilde{\u}_8
\end{bmatrix}
\cdot \begin{bmatrix}   \frac{ \| \Y_{1} \|_{\mathrm{F}}}{\sqrt{8}} & & &\\
                           & \frac{ \| \Y_{2} \|_{\mathrm{F}}}{\sqrt{8}}& & \\
                           & &  \ddots  &  \\
                           & &    &   \frac{ \| \Y_{8} \|_{\mathrm{F}}}{\sqrt{8}} \end{bmatrix}.
\end{align}
Clearly, this matrix form aligns with the one in (\ref{eq-UYG}), demonstrating that applying the cross-rotation in a similar manner will yield the $512$-dimensional reconciliation.

{Since the iterative application of the cross-rotation scheme still requires a closed-form orthogonal mapping matrix, the feasible row dimension of the newly arranged matrix are limited to $2$, $4$, or $8$. Consequently, the reconciliation dimension increases by at least a factor of two with each cross-rotation, meaning that the proposed scheme's arbitrarily high dimensions are constrained to even multiples of $8$, i.e, $d\in\{16,32,64,128,256,512,1024,\ldots\}$.}
It is also worth noting that larger cross-rotation dimensions result in increased communication overhead when transmitting rotation coefficients from Bob to Alice via the classical channel, which is on the order of $\mathcal{O}$($TN$), where $T$ represents the number of cross-rotation operations.
Along with the fact that the performance gain from increasing the rotation dimension becomes increasingly marginal (as will be shown in Fig.~\ref{fig2}), 64-dimensional reconciliation is recommended for practical implementations.


\section{Simulation and performance evaluation}

In this section, several simulations will be conducted to verify the superiority of the proposed method. We will construct a multi-edge type LDPC (MET-LDPC) code with a rate of $R_{\mathrm{code}}=0.01995$ \cite{WOS:000432546600001} and an encoded block size of $N=1\times 10^6$ bits for one-way CV-QKD reverse reconciliation, due to its high efficiency and near-Shannon limit performance with low-rate codes.
While such a one-way forward error correction using MET-LDPC has been utilized in \cite{PhysRevA.84.062317, WOS:000432546600001}, the highest available dimension of the reconciliation remains at eight due to the technical challenges mentioned above.
We will compare the proposed cross-rotation reverse reconciliation method  with the current state-of-the-art multidimensional reverse reconciliation methods for $d = 1, 2, 8$. The maximum number of iterations for MET-LDPC under sum-product decoding is set to $200$ for all strategies.
The reconciliation efficiency is defined as:
\begin{align}\label{eq-bs}
\beta(\kappa)= R_{\mathrm{code}} / \mathcal{C}(\kappa),
\end{align}
where $\mathcal{C}(\kappa)=0.5\log_2(1+ \kappa )$ is Shannon limit for a particular SNR  $\kappa$.
Equivalently, for a particular $\beta$, the SNR can be calculated as:
\begin{align}\label{eq-kappa}
\kappa(\beta)= 2^{2R_{\mathrm{code}}/\beta}-1.
\end{align}

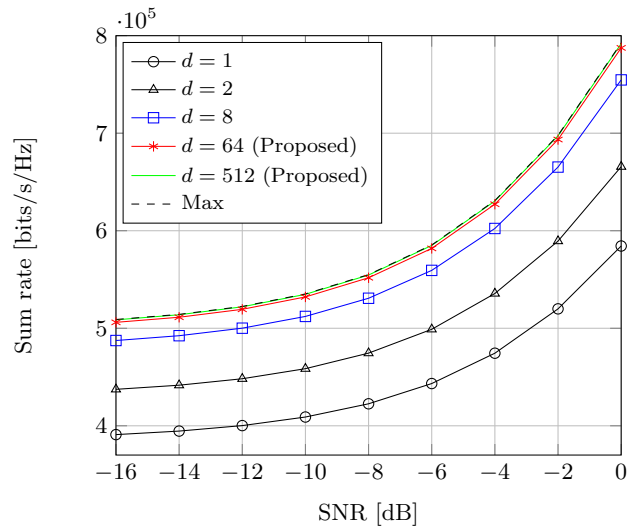
\begin{figure}
\center
\begin{tikzpicture}[scale=0.98]
\begin{axis}[xlabel={SNR [dB]},
ylabel={Sum rate [bits/s/Hz]},grid=major,
legend style={at={(0.27,0.985),font=\footnotesize}, 
anchor=north,legend columns=1,legend cell align={left}}, xmin=-16,xmax=0,ymax=800000,ymin=370000]
\addplot[mark=o,black]  coordinates{
  (    -16,  390963.644577060)
  (    -14,  394583.902582835)
  (    -12,  400210.501233070)
  (    -10,  408993.200640828)
  (     -8,  422561.801223739)
  (     -6,  443314.264159834)
  (     -4,  474428.612193868)
  (     -2,  519899.674465635)
  (      0,  584336.267552003)
};
\addplot[mark=triangle,black]  coordinates{
  (    -16, 437399.748038721)
  (    -14, 441633.267091992)
  (    -12, 448243.126772481)
  (    -10, 458562.365844287)
  (     -8, 474517.142692601)
  (     -6, 498940.394478411)
  (     -4, 535640.530479258)
  (     -2, 589390.001383331)
  (      0, 665693.183184779)
};
\addplot[mark=square,blue]  coordinates{
  (    -16,  487437.631487355)
  (    -14,  492375.135241219)
  (    -12,  500090.123046762)
  (    -10,  512137.373669538)
  (     -8,  530778.951972143)
  (     -6,  559349.085302203)
  (     -4,  602319.101761934)
  (     -2,  665278.053115382)
  (      0,  754646.757415282)
};
\addplot[mark=asterisk,red]  coordinates{
  (    -16,  506090.852799584)
  (    -14,  511296.842278091)
  (    -12,  519430.483650337)
  (    -10,  532129.683498723)
  (     -8,  551775.325325812)
  (     -6,  581889.154830261)
  (     -4,  627181.044007176)
  (     -2,  693488.867472866)
  (      0,  787463.874647307)
};

\addplot[green]  coordinates{
  (    -16,  508618.990723180)
  (    -14,  513860.099304157)
  (    -12,  521985.898250796)
  (    -10,  534806.069667763)
  (     -8,  554601.768191036)
  (     -6,  584912.277903156)
  (     -4,  630469.401275091)
  (     -2,  697301.367721079)
  (      0,  791804.327321109)
};
\addplot[dashed,black]  coordinates{
  (    -16,  508987.731058103)
  (    -14,  514233.160773120)
  (    -12,  522368.694920660)
  (    -10,  535201.201287467)
  (     -8,  555016.263596163)
  (     -6,  585359.319447755)
  (     -4,  630951.905099319)
  (     -2,  697861.195129986)
  (      0,  792442.275239728)
};
\legend{$d=1$, $d=2$, $d=8$ , $d=64$ (Proposed),  $d=512$ (Proposed),$\mathrm{Max}$}
\end{axis}
\end{tikzpicture}
\caption{Achievable sum-rate obtained from multidimensional
strategies through quantum channel versus SNR with $d=1,2,8, 64$, and $512$. The upper
bound curve (labeled \lq\lq Max") results from setting the reconciliation dimension to $N$.
 }\label{fig2}
\end{figure}

First, Monte Carlo trials are conducted to investigate the impacts of SNR (or equivalently, noise level) and reconciliation dimensions on the achievable sum-rate through the quantum channel.
Fig.~\ref{fig2} shows the achievable sum-rate obtained from different multidimensional strategies versus SNR, with results averaged over $500$ Monte Carlo quantum channel realizations, where the SNR ranges from $-16$~dB to $0$~dB.
Note that the upper bound curve (labeled \lq\lq Max") results from setting the reconciliation dimension to $N$.
We observe that: 1) the reconciliation dimension for all strategies increases with rising SNR (or equivalently, decreasing noise level); 2) increasing the reconciliation dimension from a small value to a larger value significantly enhances performance, regardless of the SNR; 3) the proposed strategy achieves a better sum-rate than the others, consistent with the theoretical analysis presented in Section 2.3; and 4) the performance gain from increasing the rotation dimension beyond $d = 64$ becomes increasingly marginal, as the $64$-dimensional reconciliation nearly approaches the upper bound.

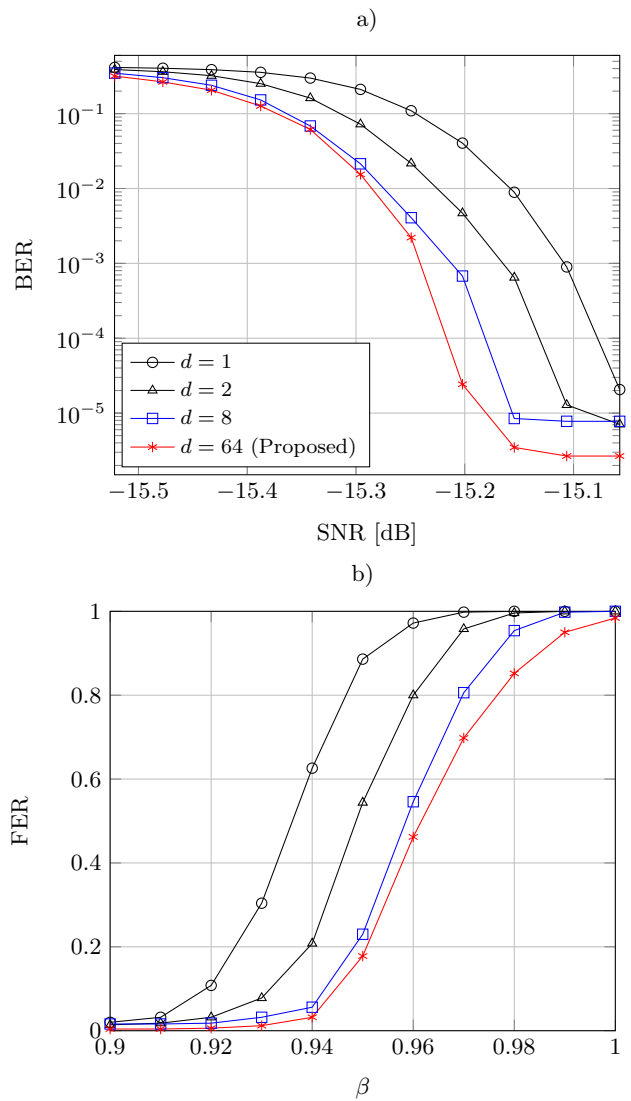
\begin{figure}
\center
\begin{tikzpicture}[scale=0.98]
\begin{semilogyaxis}[xlabel={SNR [dB]},title={a)},
ylabel={BER},grid=major,
legend style={at={(0.26,0.315),font=\footnotesize}, legend cell align={left},
anchor=north,legend columns=1},  xmin=-15.522,xmax=-15.0575,ymax=0.6,ymin=0.0000015
]
\addplot[mark=o,black]  coordinates{
  ( -15.057542335259679,   0.000020618000000)
  ( -15.106268177127358,   0.000896582000000)
  ( -15.154453478467197,   0.008909240000000)
  ( -15.202110099093321,   0.040405418000000)
  ( -15.249249512786276,   0.109376024000000)
  ( -15.295882823865135,   0.210909808000000)
  ( -15.342020782879464,   0.298721546000000)
  ( -15.387673801477186,   0.356237862000000)
  ( -15.432851966499818,   0.386906754000000)
  ( -15.477565053351984,   0.403996912000000)
  ( -15.521822538690792,   0.414220112000000)
};
\addplot[mark=triangle,black]  coordinates{
  ( -15.057542335259679,   0.000007076000000)
  ( -15.106268177127358,   0.000012862000000)
  ( -15.154453478467197,   0.000646756000000)
  ( -15.202110099093321,   0.004688206000000)
  ( -15.249249512786276,   0.021679366000000)
  ( -15.295882823865135,   0.072179484000000)
  ( -15.342020782879464,   0.161515788000000)
  ( -15.387673801477186,   0.251602340000000)
  ( -15.432851966499818,   0.321390442000000)
  ( -15.477565053351984,   0.362945234000000)
  ( -15.521822538690792,   0.388378764000000)
};
\addplot[mark=square,blue]  coordinates{
  ( -15.057542335259679,   0.000007768000000)
  ( -15.106268177127358,   0.000007768000000)
  ( -15.154453478467197,   0.000008466000000)
  ( -15.202110099093321,   0.000677000000000)
  ( -15.249249512786276,   0.004071184000000)
  ( -15.295882823865135,   0.021379356000000)
  ( -15.342020782879464,   0.068423042000000)
  ( -15.387673801477186,   0.152471500000000)
  ( -15.432851966499818,   0.239063740000000)
  ( -15.477565053351984,   0.302433238000000)
  ( -15.521822538690792,   0.347127610000000)
};
\addplot[mark=asterisk,red]  coordinates{
  ( -15.057542335259679,  0.000002668000000)
  ( -15.106268177127358,  0.000002668000000)
  ( -15.154453478467197,  0.000003480000000)
  ( -15.202110099093321,  0.000024272000000)
  ( -15.249249512786276,  0.002208532000000)
  ( -15.295882823865135,  0.015436672000000)
  ( -15.342020782879464,  0.061504458000000)
  ( -15.387673801477186,  0.126559506000000)
  ( -15.432851966499818,  0.205559596000000)
  ( -15.477565053351984,  0.265613780000000)
  ( -15.521822538690792,  0.317514634000000)
};
\legend{$d=1$, $d=2$, $d=8$ , $d=64$ (Proposed)}
\end{semilogyaxis}
\end{tikzpicture}
~~
\begin{tikzpicture}[scale=0.98]
\begin{axis}[xlabel={$\beta$},title={b)},
ylabel={FER},grid=major,
legend style={at={(0.26,0.985),font=\footnotesize}, legend cell align={left},
anchor=north,legend columns=1}, xmin=0.9,xmax=1.00,ymax=1,ymin=0
]
\addplot[mark=o,black]  coordinates{
  (     0.9000,  0.0200000000000000)
  (     0.9100,  0.0320000000000000)
  (     0.9200,  0.108000000000000)
  (     0.9300,  0.304000000000000)
  (     0.9400,  0.626000000000000)
  (     0.9500,  0.886000000000000)
  (     0.9600,  0.972000000000000)
  (     0.9700,  0.998000000000000)
  (     0.9800,  1)
  (     0.9900,  1)
  (     1.0000,  1)
};
\addplot[mark=triangle,black]  coordinates{
  (    0.9000,  0.0140000000000000)
  (    0.9100,  0.0180000000000000)
  (    0.9200,  0.0320000000000000)
  (    0.9300,  0.0780000000000000)
  (    0.9400,  0.208000000000000)
  (    0.9500,  0.544000000000000)
  (    0.9600,  0.800000000000000)
  (    0.9700,  0.958000000000000)
  (    0.9800,  0.996000000000000)
  (    0.9900,  1)
  (    1.0000,  1)
};
\addplot[mark=square,blue]  coordinates{
  (    0.9000,  0.0160000000000000)
  (    0.9100,  0.0160000000000000)
  (    0.9200,  0.0180000000000000)
  (    0.9300,  0.0320000000000000)
  (    0.9400,  0.0560000000000000)
  (    0.9500,  0.230000000000000)
  (    0.9600,  0.546000000000000)
  (    0.9700,  0.806000000000000)
  (    0.9800,  0.954000000000000)
  (    0.9900,  0.998000000000000)
  (    1.0000,  1)
};
\addplot[mark=asterisk,red]  coordinates{
  (    0.9000,   0.004000000000000)
  (    0.9100,   0.004000000000000)
  (    0.9200,   0.006000000000000)
  (    0.9300,   0.012000000000000)
  (    0.9400,   0.032000000000000)
  (    0.9500,   0.178000000000000)
  (    0.9600,   0.462000000000000)
  (    0.9700,   0.698000000000000)
  (    0.9800,   0.852000000000000)
  (    0.9900,   0.950000000000000)
  (    1.0000,   0.984000000000000)
};
\end{axis}
\end{tikzpicture}
\caption{BER and FER for a fixed MET-LDPC code with a rate of $R_{\mathrm{code}}=0.01995$. a) BER versus SNR; and b) FER versus $\beta$.
 }\label{fig3}
\end{figure}

Second, the bit error rate (BER) and frame error rate (FER) performance is evaluated for different multidimensional reconciliation schemes at low SNRs (or hight reconciliation efficiencies), where BER and FER are defined as:
\begin{align}
\mathrm{BER}= \frac{1}{NF}\sum_{f=1}^F  N_e^f
\end{align}
and
\begin{align}
\mathrm{FER}= \frac{F_e}{F},
\end{align}
where $N_e^f$ denotes the number of bit errors in the $f$-th frame, $F=500$ denotes the total number of frames, and $F_e$ denotes the number of frames in error.
As shown in (\ref{eq-bs}) and (\ref{eq-kappa}), the values of SNR and reconciliation efficiency $\beta$ correspond one-to-one. Table~\ref{tab1} presents the corresponding values of SNR and $\beta$ for a fixed MET-LDPC code rate of $R_{\mathrm{code}}=0.01995$.
%
Fig.~\ref{fig3}-a) shows the BER obtained from different multidimensional strategies versus SNR under soft-decision sum-product decoding. It is observed that: 1) the BER performance for all strategies improves with increasing SNR; and 2) the proposed method consistently achieves the best BER performance for any fixed SNR, as its higher reconciliation dimension results in the greatest performance improvement.
Fig.~\ref{fig3}-b) shows the FER obtained from different multidimensional strategies versus reconciliation efficiency $\beta$ under soft-decision sum-product decoding. It is observed that: 1) $8$-dimensional reconciliation exhibits significantly better FER performance than $1$-dimensional and $2$-dimensional reconciliations, with its FER attaining 1 when $\beta= 0.99$; and 2) the proposed method consistently outperforms $8$-dimensional reconciliation, with the maximum available $\beta$ reaching up to $1$.
Considering that the codes are designed to operate with high $\beta$ efficiency (even at a high FER) to maximize the secret key rate and distance, the proposed method demonstrates greater practical potential for long-distance CV-QKD.

\begin{table}
\center
\caption{Corresponding values of SNR and $\beta$ for a fixed MET-LDPC code rate of $R_{\mathrm{code}}=0.01995$.}\label{tab1}
\begin{tabular}{c|cccc}
  \hline
  SNR [dB] & $ -15.0575$ & $-15.1063$ & $-15.1545$ & $-15.2021$  \\
  $\beta$  & $0.90     $ & $0.91    $ & $0.92    $ & $0.93    $  \\
  \hline
  \hline
  SNR [dB] & $-15.2492$ & $-15.2959$ & $-15.3420$  & $-15.3877$ \\
  $\beta$  & $0.94    $ & $0.95    $ & $0.96    $  & $0.97    $   \\
  \hline
  \hline
  SNR [dB] & $-15.4329$ & $-15.4776$ & $-15.5218$ & \\
  $\beta$  & $0.98    $ & $0.99    $ & $1.00    $ &   \\
  \hline

\end{tabular}
\end{table}

\begin{figure}
\center
\begin{tikzpicture}[scale=0.98]
\begin{axis}[xlabel={$\beta$},title={a)},
ylabel={Number of iterations},grid=major,
legend style={at={(0.7399,0.315),font=\footnotesize}, legend cell align={left},
anchor=north,legend columns=1},  xmin=0.9,xmax=1.00 ,ymax=205,ymin=60
]
\addplot[mark=o,black]  coordinates{
  (  0.9000,   97.1660000000000)
  (  0.9100,   113.510000000000)
  (  0.9200,   136.524000000000)
  (  0.9300,   163.190000000000)
  (  0.9400,   185.170000000000)
  (  0.9500,   196.422000000000)
  (  0.9600,   199.438000000000)
  (  0.9700,   199.970000000000)
  (  0.9800,   200)
  (  0.9900,   200)
  (  1.0000,   200)
};
\addplot[mark=triangle,black]  coordinates{
  (  0.9000,  80.6480000000000)
  (  0.9100,  91.1280000000000)
  (  0.9200,  106.106000000000)
  (  0.9300,  126)
  (  0.9400,  152.744000000000)
  (  0.9500,  178.556000000000)
  (  0.9600,  193.348000000000)
  (  0.9700,  198.996000000000)
  (  0.9800,  199.886000000000)
  (  0.9900,  200)
  (  1.0000,  200)
};
\addplot[mark=square,blue]  coordinates{
  (  0.9000,  73.2040000000000)
  (  0.9100,  81.4020000000000)
  (  0.9200,  92.1820000000000)
  (  0.9300,  106.420000000000)
  (  0.9400,  126.810000000000)
  (  0.9500,  153.816000000000)
  (  0.9600,  177.988000000000)
  (  0.9700,  193.286000000000)
  (  0.9800,  198.958000000000)
  (  0.9900,  199.976000000000)
  (  1.0000,  200)
};
\addplot[mark=asterisk,red]  coordinates{
  (  0.9000, 70.7100000000000)
  (  0.9100, 78.3720000000000)
  (  0.9200, 88.2620000000000)
  (  0.9300, 101.678000000000)
  (  0.9400, 120.968000000000)
  (  0.9500, 147.340000000000)
  (  0.9600, 172.280000000000)
  (  0.9700, 186.150000000000)
  (  0.9800, 194.424000000000)
  (  0.9900, 198.234000000000)
  (  1.0000, 199.672000000000)
};
\legend{$d=1$, $d=2$, $d=8$ , $d=64$ (Proposed)}
\end{axis}
\end{tikzpicture}
~~
\begin{tikzpicture}[scale=0.98]
\begin{semilogyaxis}[xlabel={$\beta$},title={b)},
ylabel={Runtime [seconds]},grid=major,
legend style={at={(0.7399,0.315),font=\footnotesize}, legend cell align={left},
anchor=north,legend columns=1},  xmin=0.9,xmax=1.00 ,ymax=84000,ymin=26500
]
\addplot[mark=o,black]  coordinates{
  (  0.9000, 38429.1709898000)
  (  0.9100, 44811.9766818000)
  (  0.9200, 54282.3448785000)
  (  0.9300, 65303.3483838000)
  (  0.9400, 74390.6251711000)
  (  0.9500, 79314.7984145000)
  (  0.9600, 80808.9048106000)
  (  0.9700, 81284.6808856000)
  (  0.9800, 81338.5351288000)
  (  0.9900, 81418.3613042000)
  (  1.0000, 81443.1523154000)
};
\addplot[mark=triangle,black]  coordinates{
  (  0.9000, 32677.3602881000)
  (  0.9100, 36862.4649145000)
  (  0.9200, 43013.7742197000)
  (  0.9300, 51076.0462805000)
  (  0.9400, 61816.5957202000)
  (  0.9500, 72529.6474851000)
  (  0.9600, 78934.6399698000)
  (  0.9700, 81648.4220938000)
  (  0.9800, 82175.3773928000)
  (  0.9900, 82241.4089877000)
  (  1.0000, 82184.2658605000)
};
\addplot[mark=square,blue]  coordinates{
  (  0.9000, 29306.1481537000)
  (  0.9100, 32728.6139656000)
  (  0.9200, 37098.3127064000)
  (  0.9300, 42872.5312443000)
  (  0.9400, 50962.5201461000)
  (  0.9500, 61963.5523143000)
  (  0.9600, 71952.6449615000)
  (  0.9700, 78600.0620393000)
  (  0.9800, 81122.4079577000)
  (  0.9900, 81652.9423917000)
  (  1.0000, 81647.4298973000)
};
\addplot[mark=asterisk,red]  coordinates{
  (  0.9000, 29423.1917667000)
  (  0.9100, 32726.9167422000)
  (  0.9200, 36786.6674693000)
  (  0.9300, 42177.2931061000)
  (  0.9400, 49799.0673788000)
  (  0.9500, 60506.8203627000)
  (  0.9600, 70747.1390860000)
  (  0.9700, 76629.4719712000)
  (  0.9800, 80156.1059815000)
  (  0.9900, 82133.0453994000)
  (  1.0000, 82981.6101941000)
};
\end{semilogyaxis}
\end{tikzpicture}
\caption{Number of iterations and CPU runtime required by different multidimensional strategies versus $\beta$. a) number of iterations versus $\beta$; and b) runtime versus $\beta$.
 }\label{fig4}
\end{figure}
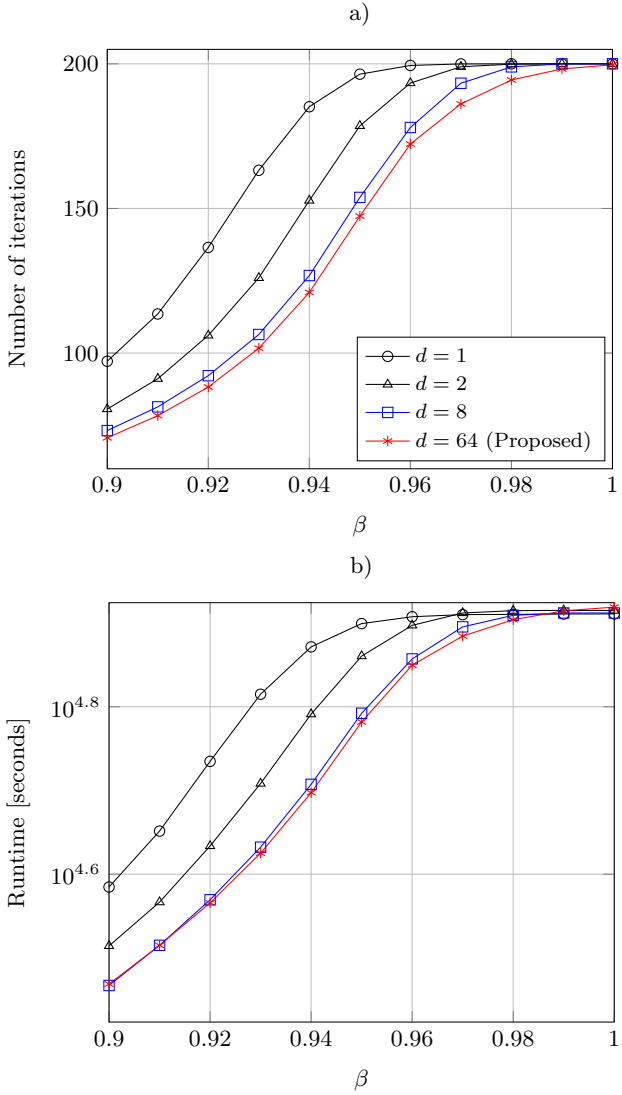

Third, we study the effect of the reconciliation dimension on the required number of iterations and the CPU runtime. Fig.~\ref{fig4}-a) shows the number of iterations required by different multidimensional strategies versus $\beta$, while Fig.~\ref{fig4}-b) illustrates the overall CPU runtime consumed by these strategies versus $\beta$. It is observed that: 1) the runtimes of all strategies are closely related to their required number of iterations; specifically, fewer iterations generally result in shorter runtimes; 2) the proposed method has a computational cost advantage despite involving more rotation operations, as it requires fewer iterations by exploiting higher dimensions; and 3) all methods exhibit similar runtimes when the $\beta$ is sufficiently large, as most cannot successfully correct errors and fail to converge.


Fourth, we investigate the impact of distance on the finite-size SKR in a CV-QKD system with one-way reverse reconciliation. We assume that half of the bits ($N=10^6$) are used for parameter estimation and the other half ($N=10^6$) are utilized in the information reconciliation process to generate secret keys. The finite-size SKR for homodyne detection is calculated as \cite{PhysRevA.81.062343}:
\begin{align}
\mathrm{SKR}=\frac{1}{2}(1-\mathrm{FER})(\beta I(A;B)- S_{\epsilon}(B;E)- \Delta(N) ),
\end{align}
where $I(A;B)$ denotes the mutual information between Alice and Bob,
$S_{\epsilon}(B;E)$ represents the Holevo bound under finite-size effect, and $ \Delta(N)$ denotes the finite-size offset factor \cite{PhysRevA.81.062343,PhysRevA.76.042305}. Both $S_{\epsilon}(B;E)$ and $\Delta(N)$ can be precisely calculated as described in \cite{Jiang2024}.
The channel transmittance $T(d)=10^{-\alpha d/10}$ is determined by the standard loss of $\alpha=0.2$ dB/km and the distance $d$ in kilometers. The excess channel noise is measured in shot-noise units as \cite{WOS:000432546600001}:
\begin{align}
\xi=\begin{cases}0.01, &  0 \le d\le 100 ~\mathrm{km}\\
                 0.01+ 0.001(d-100) ,& d > 100 ~\mathrm{km}   \end{cases}.
\end{align}
Bob's homodyne detector efficiency is set at $\eta = 0.606$ with an added electronic noise of $v_{el}=0.41$. To simplify the selection of Alice's modulation variance and ensure that the SNR remains constant across the entire distance range, we set:
\begin{align}
V_A(\beta, d)=\kappa(\beta) ( 1+\chi_{tot}(d)  ),
\end{align}
where $\kappa(\beta)$ has been defined in (\ref{eq-kappa}) and  $\chi_{\mathrm{tot}}(d)= \chi_{\mathrm{line}}(d) +  \chi_{\mathrm{hom}}/T(d)$ denotes the total noise added into the input of quantum channel with
$\chi_{\mathrm{line}}(d) = 1/T(d)  + \xi -1 $  and $\chi_{\mathrm{hom}}= (1+v_{el})/\eta  -1  $.

\begin{figure}
\center
\includegraphics[width=8.2cm]{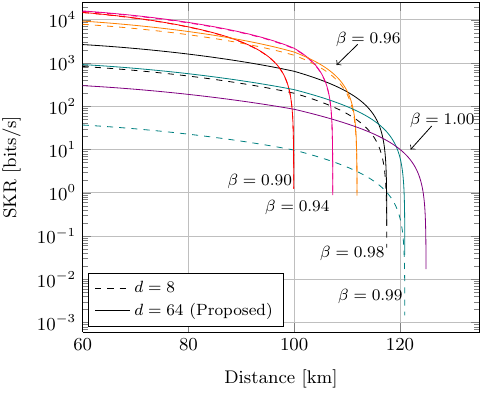}\\
%
%
%
\caption{Finite-size SKR for a one-way reverse reconciliation CV-QKD system at a repetition rate of 5 MHz over the relevant distance range.}\label{fig5}
\end{figure}

Fig.~\ref{fig5} illustrates the finite-size SKR results for a CV-QKD system utilizing one-way reverse reconciliation at a repetition rate of $5$~MHz across the relevant distance range. The following observations can be made:
1) the $8$-dimensional reconciliation and the proposed method achieve similar finite-size SKR and reconciliation distances for $\beta\le 0.94$;
2) the proposed method surpasses the $8$-dimensional reconciliation when $\beta$ exceeds $0.96$, due to its significantly better frame error rate (FER) performance, as shown in Fig.~3-b; 3) for $\beta > 0.99$, the FER becomes a limiting factor for achieving a non-zero SKR in the $8$-dimensional reconciliation, with a maximum distance of $121$ km; and 4) the proposed method operates effectively up to $\beta = 1$, achieving a maximum distance of approximately $130$~km under the finite-size effect.

\begin{table}
\center
\caption{\textcolor[rgb]{0.00,0.00,0.00}{Corresponding values of SNR and $\beta$ for a fixed ATSC 3.0 LDPC code rate of $R_{\mathrm{code}}=0.2$.}}\label{tab2}
\begin{tabular}{c|cccc}
  \hline
  SNR [dB] & $ -4.4275$ & $-4.4832$ & $-4.5382$ & $-4.5925$  \\
  $\beta$  & $0.90     $ & $0.91    $ & $0.92    $ & $0.93    $  \\
  \hline
  \hline
  SNR [dB] & $-4.6462$ & $-4.6992$ & $-4.7516$  & $-4.8034$ \\
  $\beta$  & $0.94    $ & $0.95    $ & $0.96    $  & $0.97    $   \\
  \hline
  \hline
  SNR [dB] & $-4.8546$ & $-4.9052$ & $-4.9552$ & \\
  $\beta$  & $0.98    $ & $0.99    $ & $1.00    $ &   \\
  \hline

\end{tabular}
\end{table}

\begin{figure}
\center
\begin{tikzpicture}[scale=0.98]
\begin{semilogyaxis}[xlabel={SNR [dB]},title={a)},
ylabel={BER},grid=major,
legend style={at={(0.290,0.249),font=\footnotesize}, legend cell align={left},
anchor=north,legend columns=1},  xmin=-4.95518384571360,xmax=-4.5,ymax=0.35,ymin=0.0003
]
\addplot[mark=square,blue]  coordinates{
  ( -4.42745508073754,  0                )
  ( -4.48316984930666,  0                )
  ( -4.53818818923575,  0.000660651234567)
  ( -4.59252711481114,  0.044862898148148)
  ( -4.64620302966900,  0.188903246913580)
  ( -4.69923175546274,  0.247128888888889)
  ( -4.75162855887796,  0.255213385802469)
  ( -4.80340817710713,  0.260034598765432)
  ( -4.85458484188798,  0.263841777777778)
  ( -4.90517230220084,  0.267060432098765)
  ( -4.95518384571360,  0.269896712962963)
};
\addplot[mark=asterisk,red]  coordinates{
  ( -4.42745508073754, 0)
  ( -4.48316984930666, 0)
  ( -4.53818818923575, 0)
  ( -4.59252711481114, 0)
  ( -4.64620302966900, 0)
  ( -4.69923175546274, 0.016933530864197)
  ( -4.75162855887796, 0.112004265432099)
  ( -4.80340817710713, 0.210232179012346)
  ( -4.85458484188798, 0.246192487654321)
  ( -4.90517230220084, 0.253182419753086)
  ( -4.95518384571360, 0.257717114197531)
};
\addplot[mark=pentagon,black]  coordinates{
  ( -4.42745508073754,  0)
  ( -4.48316984930666,  0)
  ( -4.53818818923575,  0)
  ( -4.59252711481114,  0)
  ( -4.64620302966900,  0.000405074074074)
  ( -4.69923175546274,  0.016792314814814)
  ( -4.75162855887796,  0.110430111111111)
  ( -4.80340817710713,  0.208519870370370)
  ( -4.85458484188798,  0.245572364197531)
  ( -4.90517230220084,  0.253198765432099)
  ( -4.95518384571360,  0.257751709876543)
};
\legend{$d=8$,$d=64$ (Proposed), $d=64$ (Householder)}
\end{semilogyaxis}
\end{tikzpicture}
~~
\begin{tikzpicture}[scale=0.98]
\begin{axis}[xlabel={$\beta$},title={b)},
ylabel={FER},grid=major,
legend style={at={(0.285,0.965),font=\footnotesize}, legend cell align={left},
anchor=north,legend columns=1}, xmin=0.9,xmax=1.00,ymax=1,ymin=0
]
\addplot[mark=square,blue]  coordinates{
  (    0.9000, 0    )
  (    0.9100, 0    )
  (    0.9200, 0.004)
  (    0.9300, 0.366)
  (    0.9400, 0.970)
  (    0.9500, 1    )
  (    0.9600, 1    )
  (    0.9700, 1    )
  (    0.9800, 1    )
  (    0.9900, 1    )
  (    1.0000, 1    )
};
\addplot[mark=asterisk,red]  coordinates{
  (    0.9000,  0    )
  (    0.9100,  0    )
  (    0.9200,  0    )
  (    0.9300,  0    )
  (    0.9400,  0    )
  (    0.9500,  0.178)
  (    0.9600,  0.868)
  (    0.9700,  0.998)
  (    0.9800,  1    )
  (    0.9900,  1    )
  (    1.0000,  1    )
};
\addplot[mark=pentagon,black]  coordinates{
  (    0.9000,  0    )
  (    0.9100,  0    )
  (    0.9200,  0    )
  (    0.9300,  0    )
  (    0.9400,  0.004)
  (    0.9500,  0.172)
  (    0.9600,  0.868)
  (    0.9700,  1    )
  (    0.9800,  1    )
  (    0.9900,  1    )
  (    1.0000,  1    )
};
\end{axis}
\end{tikzpicture}
\caption{BER and FER for a fixed ATSC 3.0 LDPC code rate of $R_{\mathrm{code}}=0.2$. a) BER versus SNR; and b) FER versus $\beta$.}\label{fig6}
\end{figure}
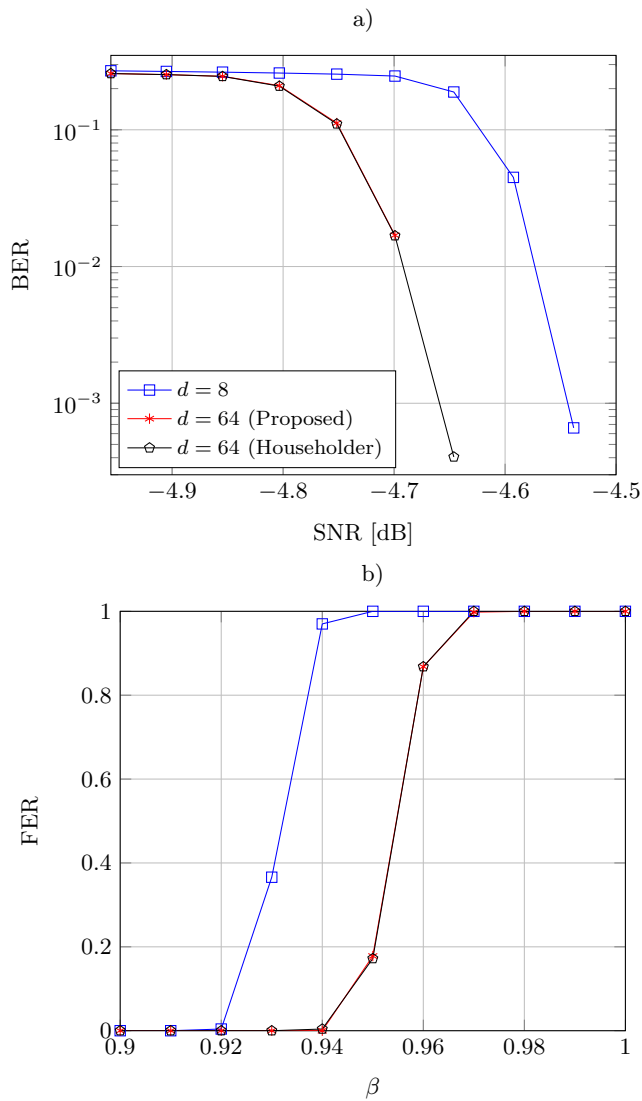

{Finally, we evaluate the performance of the proposed method, the original multidimensional method, and the Householder-based method for a relatively high LDPC code rate. An ATSC 3.0 LDPC code \cite{e22101087} with a rate of $R_{\mathrm{code}}=0.2$ and an encoded block size of $N=648000$  bits is constructed for one-way CV-QKD reverse reconciliation.
Fig.~\ref{fig6} shows the BER and FER obtained from different multidimensional strategies
versus SNR and  reconciliation efficiency $\beta$, respectively.
Table~\ref{tab2} presents the corresponding values of SNR and $\beta$ for a fixed ATSC 3.0 LDPC code rate of $R_{\mathrm{code}}=0.2$.
It is observed that the proposed method and the Householder-based method achieve nearly identical BER and FER performance, consistently outperforming the original multidimensional reconciliation ($d=8$). The reason for this is that both methods use the same dimension ($d=64$). These results align with our performance analysis in Section III-D, which confirms that the achievable sum-rate is directly determined by the reconciliation dimension.
Notably, since the communication overhead for transmitting a $64$-dimensional mapping matrix is $64^2$, the total communication overhead for the Householder-based $64$-dimensional reconciliation is $64N$, which is significantly higher than that of the proposed method ($2N$).}

\section{Conclusion}

This paper introduces a cross-rotation scheme that addresses the limitation of 8-dimensional rotations in multidimensional reconciliation, extending the process to higher dimensions. By reshaping the string vector into matrix form and applying orthogonal transformations to both columns and rows, the reconciliation dimension is increased by at least one order per cross-rotation. A rigorous performance analysis is provided, and simulations confirm the effectiveness of the proposed approach. While the method allows for extending the reconciliation dimension arbitrarily further, practical considerations such as increased communication traffic and diminishing performance (as shown in Fig.~\ref{fig2}) returns suggest that $64$-dimensional reconciliation strikes the best balance for practical implementations.

%

\appendix

\section{Majorization}

We  introduce some fundamental definitions and a lemma related to majorization that will be used for performance analysis in Section III-D. For a detailed explanation of majorization, readers can refer to \cite{marshall11, Palomar2007}.

\textbf{Definition 1}. For any $\a,\b\in \mathbb{R}^{Q}$, we say that  $\a$ is majorized by $\b$ (denoted as $\a \prec \b$) if the following conditions hold:
\begin{align}
&\sum_{q=1}^P a_{[q]}  \le \sum_{q=1}^P b_{[q]}, ~1\le P\le Q-1,\\
&\sum_{q=1}^Q a_{[q]}=\sum_{q=1}^Q b_{[q]},
\end{align}
where $a_{[1]}\ge a_{[2]}\ge \ldots \ge a_{[Q]} $ represents the components of $\a$ arranged in non-increasing order, and similarly, $b_{[1]}\ge b_{[2]}\ge \ldots \ge b_{[Q]} $ for the non-increasing  components of $\b$.

\textbf{Definition 2}. For any $\a, \b \in \mathcal{A} \subseteq \mathbb{R}^{Q}$, a real-valued function $\varphi$ is said to be Schur-convex on $\mathcal{A}$ if
\begin{align}
\varphi(\a) \le \varphi(\b),  ~\mathrm{whenever}~\a \prec \b.
\end{align}
While $\varphi$ is said to be Schur-concave on $\mathcal{A}$ if
\begin{align}
\varphi(\a) \ge \varphi(\b),  ~\mathrm{whenever}~\a \prec \b.
\end{align}

\textbf{Lemma 3}. $f(\b)= \sum_{q=1}^{Q}\log_2(1+  b_q)  $ is  Schur-concave on the set of $\{\b| b_q\ge 0, q=1,2,\ldots, Q\}$, where $b_q$ denotes the $q$-th element of $\b$.
\begin{proof}
The proof follows directly from Lemma 1 in \cite{Dai2014}. For completeness, we provide a brief sketch below.
 \begin{itemize}
  \item It is straightforward to prove $ f(\b) \le f(\T\b)$, where $\T$ is any $T$-transform matrix of the form $\lambda\I + (1-\lambda)\P$ with $0\le\lambda\le1$ and $\P$ being a permutation matrix that interchanges two coordinates.
  \item For $\a \prec \b$, $\a$ can be derived from $\b$ by successive
applications of a finite number of $T$-transforms (see 2.B.1. Lemma in \cite{marshall11}).
\end{itemize}
Therefore, we obtain the desired result $f(\b) \leq f(\a)$.
\end{proof}

\section{Mutual information for statistical independence variables}
The mutual information between $\mathcal{X}$ and $\mathcal{Y}$ is given by \cite{Cover2006}:
\begin{align}
I(\mathcal{X};\mathcal{Y})= \int p(x,y)\log\frac{p(x,y)}{p(x)p(y)}  dxdy.
\end{align}
If  $\mathcal{X}$ and $\mathcal{Y}$ are statistically independent, i.e., $p(y|x)=p(y)$, the mutual information $I(\mathcal{X};\mathcal{Y})$ is zero, because
\begin{align}
I(\mathcal{X};\mathcal{Y}) =& \int p(x,y)\log\frac{p(y|x)}{p(y)}dxdy \notag\\
=& \int p(x,y)\log\frac{p(y)}{p(y)} dxdy=0.
\end{align}

\bibliography{apssamp}

\end{document}